\documentclass[12pt]{article}

\usepackage{graphicx}
\usepackage{amsmath,amssymb,amsfonts}
\usepackage{amsthm}
\usepackage{geometry}
\usepackage{caption}
\usepackage{subcaption}
\usepackage{float}
\usepackage{multirow}
\usepackage{booktabs}
\usepackage{xcolor}
\usepackage{textcomp}
\usepackage{url}
\usepackage{natbib}
\usepackage{hyperref}

\hypersetup{
    colorlinks=true,
    linkcolor=blue,
    citecolor=blue,
    urlcolor=blue
}

\theoremstyle{plain}

\theoremstyle{definition}

\usepackage{setspace}

\begin{document}

\title{Mapping the Midweek Mountain: \\ The New Geography of Hybrid Work}

\author{
Norman Guo\footnote{Saint Louis University. Email: norman.guo@slu.edu} \and 
Wei Jiang\footnote{Corresponding author. Emory University, NBER, and ECGI. Email: wei.jiang@emory.edu} \and 
Yaswanth Pothuru\footnote{Georgia State University. Email: ypothuru1@gsu.edu} \and 
Baozhong Yang\footnote{Georgia State University. Email: bzyang@gsu.edu}
}

\date{March 2026}

\maketitle

\begin{abstract}
This paper provides a behavioral analysis of the post-pandemic transformation of work, using a dataset of approximately 41 billion mobile geolocation records from 73.5 million individuals in the five largest U.S. metropolitan areas from the pre- to post-pandemic periods. By tracking movements between corporate headquarters, residences, and other points of interest, we document a structural shift in work patterns. Office-based workdays declined from 42\% in 2019 to 20.7\% in 2022, before settling at 29.1\% in 2023, a new equilibrium significantly below pre-pandemic levels. A ``midweek mountain'' peak of office attendance on Tuesdays through Thursdays, emerged as a robust new phenomenon post-pandemic. The nature of remote work has also changed: both in and after the pandemic, employees working from home allocated significantly more time to non-work locations like parks and malls during the workday. These findings indicate that the pandemic catalyzed a lasting transformation not just in work arrangements but also in the integration of personal and professional life, with implications for corporate policy, urban economics, and the future of work.
\end{abstract}


\medskip
\noindent\textbf{Classification:} Social Sciences, Economic Sciences

\newpage

\section*{\centering Significance Statement}

\setstretch{1.5}

\bigskip


\noindent Most evidence on remote work relies on surveys and self-reported data. This study uses billions of anonymized mobile geolocation records to directly observe how workers move among offices, homes, and other locations during the workday. We document three central findings. First, the pandemic triggered a lasting reduction in office attendance that has not reversed. Second, a distinct ``midweek mountain" has emerged, with in-person work concentrated on Tuesdays through Thursdays. Third, and most novel, workers now spend significantly more time at leisure and amenity locations such as parks and malls during work hours, regardless of whether they commute to the office. This behavioral evidence reveals a fundamental rebalancing of professional and personal activities, with broad implications for organizational policy, urban planning, and the future of work.

\newpage


\section{Introduction}\label{sec1}
The Covid-19 pandemic triggered a large-scale, unplanned experiment in working from home (WFH). As offices closed and mobility restrictions took effect, remote work became the default mode for many knowledge workers worldwide. While prior studies have documented the prevalence of remote work using surveys and self-reported data,\footnote{For examples, see \cite{aksoy2022, brynjolfsson2020, barrero2021, bloom2024}.} there remains limited evidence on the actual large-sample behavioral patterns and spatial dynamics of how employees navigate between home, office, and other locations during and after the pandemic. In this paper, we address this gap by leveraging a novel dataset of anonymized geolocation pings to provide empirical evidence on how work location patterns evolved across the pre-pandemic, pandemic, and post-pandemic periods (2019--2023).

In this study, we analyze a dataset containing approximately 41 billion geolocation records from nearly 73.5 million mobile devices across the five largest Metropolitan Statistical Areas (MSAs) in the United States. The dataset covers more than 10\% of the U.S. population, offering the highest level of coverage currently available \citep{hsu2024}. Using a systematic procedure that links pings from the devices to corporate headquarters, residential locations, and points of interest, we identify employees of publicly traded firms with well-defined addresses in these MSAs and track their movements among home, office, and other locations during work hours.

Our analysis confirms a structural shift away from the office. We find that office-based workdays plummeted from 42.0\% in 2019 to 20.7\% in 2022, recovering only to 29.1\% in 2023, suggesting a new equilibrium significantly below pre-pandemic levels instead of a full recovery. We also document substantial heterogeneity across sectors and geographies. Industries like Information Technology, Consumer Staples, and Finance exhibit the most profound transformations, while geographically, MSAs such as San Francisco and New York show a more enduring adoption of remote work compared to Chicago.

Crucially, our granular data reveals novel insights into the \textit{intensity} and \textit{timing} of office work that challenge the binary view of ``in-office'' versus ``remote'' days. We find that office attendance is declining not just in frequency, but also in duration. Even on days when employees commute to the office, the average time spent physically on-site decreased by approximately 36 minutes between 2019 and 2023 (from 442.1 minutes to 406.5 minutes out of a standard 480-minute workday). This suggests that the traditional expectation of full-day, in-person presence is eroding. Furthermore, a distinct weekday pattern has stabilized, characterized by a \textit{midweek mountain} where office presence peaks Tuesday through Thursday and drops significantly on Mondays and Fridays, confirming the adoption of coordinated hybrid schedules.\footnote{This pattern aligns with recent survey evidence from industry reports documenting national trends in office occupancy \citep{kastle2023, gallup2023}. Prior academic research based on surveys and self-reported data also note that the typical hybrid arrangement involves commuting to the office mostly from Tuesdays to Thursdays \citep{barrero2023, choudhury2024, bloom2024}.} 
The emergence of this new regularity likely reflects both employer coordination to maximize in-person collaboration and employees' preferences for balancing office presence with remote flexibility.

The granular nature of our data allows us to analyze detailed behavior patterns that are beyond the scope of typical surveys and reports. In particular, we are able to study how employees utilize the flexibility afforded by remote work. In contrast to the pre-pandemic era, where working from home largely meant staying in place, we find that in 2022-2023, individuals allocated significantly more time to non-work locations such as parks, malls, and golf clubs during standard work hours. Strikingly, such behavior remains largely unchanged after the pandemic, and are persistent whether or not employees work from home or commute to offices. This provides direct empirical evidence of a profound blurring of boundaries between workplace and broad activities. The flexibility of remote work is characterized not just by a change of venue, but by a fundamental rebalancing of office and broad-venue activities.

These findings deepen our understanding of how the pandemic could have uprooted traditional work patterns such that the shift to remote and hybrid work reflects a structural transformation rather than a temporary adjustment. The emergence of the pronounced “midweek mountain” in office attendance suggests a natural coordination point for in-person work that may evolve into a more concentrated—and potentially shorter—workweek. This possibility aligns with a growing discussion that advances in digital technologies and artificial intelligence prowess can sustain output with fewer labor hours (e.g., \citealt{brynjolfsson2014,agrawal2019}). By providing granular, behavior-based evidence, our study offers implications for organizational design, urban planning, and post-pandemic economic policy.

Our study builds on an extensive body of research examining the transformation of work catalyzed by the COVID-19 pandemic. While early research on telework established that remote arrangements could offer benefits like increased job satisfaction without compromising performance \citep{gajendran2007}, the pandemic triggered an unprecedented, large-scale shift. Foundational studies by \citet{bloom2024} and \citet{barrero2021} used large-scale surveys and randomized controlled trials to document this change, showing that remote work spiked from roughly 4\% of U.S. workdays pre-pandemic to a stable post-pandemic norm of around 25\%. Their work established that this shift is persistent, with both employees and employers embracing hybrid models, and that such arrangements can improve employee retention significantly.

Subsequent research has explored the multifaceted impacts of this transformation. While hybrid work has been linked to higher job satisfaction, studies also highlight significant challenges, particularly for collaboration and innovation. For instance, \citet{yang2022} found that firm-wide remote work caused communication to become more static and siloed, hindering the cross-functional knowledge sharing that often sparks new ideas. Beyond the firm, the shift to remote work has generated profound economic spillovers, reshaping housing markets as employees are no longer tethered to urban centers \citep{mondragon2022} and creating significant disruption in the commercial real estate sector \citep{wirz2025}.

While these studies provide crucial insights, much of the evidence is derived from surveys, which rely on self-reported data. A newer stream of research has begun to leverage high-frequency geomobile data to capture actual, observed behavior. Early work during the pandemic used mobility data to document the initial, dramatic changes in commuting and consumption patterns \citep{brynjolfsson2020}. Methodological advances have also shown that smartphone data can reliably infer home and work locations, revealing complex travel and activity patterns that surveys cannot capture \citep{miyauchi2021}. Our paper contributes to this emerging literature by using a massive, granular geolocation dataset to provide a behavioral analysis of how work patterns, including daily routines and time allocation, have structurally evolved from the pre-pandemic era to the new post-pandemic equilibrium.

\section{Data Description and Processing}\label{sec3}

Key to this study is data from Veraset, a data broker that sells large-scale, pseudonymous smartphone location datasets via raw GPS pings and inferred visits/trips. Given our research interest as well as for tractability, we focus on the top five Metropolitan Statistical Areas (MSAs) in the United States: New York, Los Angeles, Chicago, Miami, and San Francisco. For each of these regions, we collect anonymized geolocation pings and associated timestamps for 2019 (``pre-pandemic'') and 2022–2023 (``post-pandemic''). The raw dataset contains approximately 41 billion pings over these three years, associated with nearly 73.5 million mobile devices.\footnote{In the dataset, a mobile device is considered a unique, including a new, device after each update of its operating system and assigned a new id in the dataset.} These devices could belong to any parts of the population including residents and visitors at the particular MSAs. 

Each data point, referred to as a ``ping,'' represents a discrete spatiotemporal record comprising a timestamp and geolocation (latitude and longitude), all associated with a unique mobile device identifier. These pings are generated from third-party mobile applications (e.g., weather, navigation, or utility apps) that utilize the device’s location services framework. The temporal frequency of these pings is not always continuous, with generation triggered by user interaction with apps and background processes. Consequently, the data exhibit naturally heterogeneous granularity, where the typical update intervals range from seconds to minutes.

Given our interest in analyzing working patterns, we restrict the sample to employees who are local residents. We further focus on publicly traded firms, whose headquarters can be precisely located using disclosure filings and which typically occupy independent, identifiable buildings. These characteristics are essential for distinguishing employment-related presence from non-employment visits. Because a central objective of the study is to examine the fluidity between work-from-office and work-from-home arrangements, we develop algorithms to identify individuals employed by public companies whose workplace and residential locations fall within the same metropolitan area.

To identify workplaces, we retrieve headquarters' address information from firms' annual 10-K SEC filings and cross-reference these with shapefiles of building polygons for each state to construct precise headquarters boundaries. A geolocation ping falling within a headquarter's polygon is classified as an office visit. To ensure that we capture employees instead of visitors to a company, we impose additional filters. First, we retain only individuals with observation data spanning at least 50 distinct work days over the sample period to allow sufficient data. Work days are defined to be 9am to 5pm from Monday to Friday, excluding federal holidays. Second, we require that at least 20\% of these observed work days include office visits to a given company. When an individual visits multiple headquarters, we assign the person to the company most frequently visited.

To infer home locations, we analyze geolocation pings recorded between 12am and 5am. The location with the highest number of nighttime pings is labeled as the individual's home, conditional on the location being visited during at least 50 distinct days every year. This threshold is designed to guarantee the stability and persistence of the home location inference, avoiding noise from short-term or transient stays. For other location visits, we match geolocation pings to SafeGraph's Points of Interest (POI) database, which is a comprehensive dataset covering millions of businesses, parks, apartments, and other categories of buildings and structures in the US.


Our final sample consists of 3,237 employees in 2019, 6,338 employees in 2022, and 3,092 employees in 2023, affiliated with 482, 621, and 643 companies, respectively. For each individual, we calculate the total time spent at various locations (office, home, or other points of interest) during work days (excluding weekends and Federal holidays) between 9am and 5pm. The time spent at a location is calculated from the first ping at the location to the first ping at the next location. The total time of each individual spent at all locations in a workday is scaled to a standardized eight-hour workday, or 480 minutes. In our baseline classification, we label a person–day as \textit{work-from-office} (WFO) if, within the standard eight-hour work window, the individual spends at least 120 minutes at the office and this office time exceeds the time spent at any other single location on that day. Analogously, we classify a person–day as \textit{work-from-home} (WFH) if the individual spends at least 120 minutes at home during the work window and this home time is greater than the time spent at any other location on that day.\footnote{Although individuals are also observed to work predominantly from other locations, this study focuses on days characterized by either home-based or office-based work. By doing this, we enhance the comparability of home and office work patterns across individuals and over time. Our main results are qualitatively robust to alternative choices of the 120-minute threshold in the definitions of WFH and WFO days.}



\section{Main Findings}\label{sec4}

\subsection{Workday allocation at office and home}

\begin{figure}[H]
\centering
\includegraphics[width=0.9\textwidth]{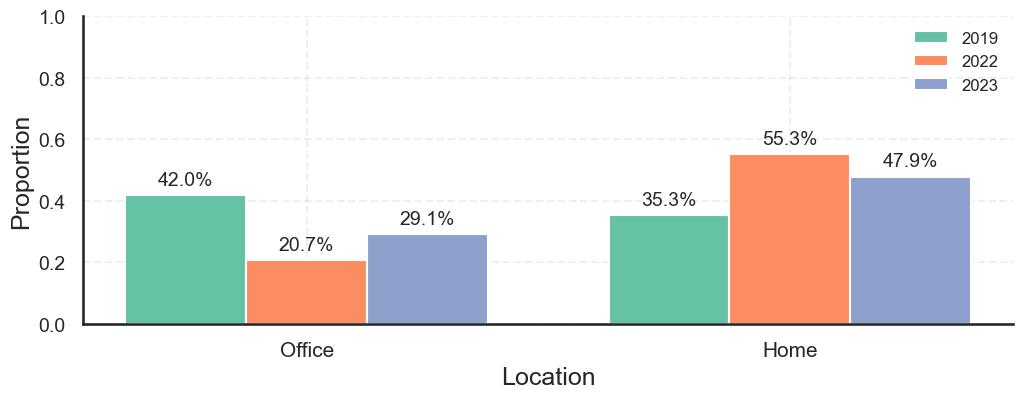}
\caption{\textbf{Shifts in Work Location Preferences:} Proportion of Workdays Spent in the Office vs. at Home in 2019, 2022, and 2023}\label{fig1}
\end{figure}

The proportion of work days spent at the office and at home shows substantial variation across 2019, 2022, and 2023 (Figure \ref{fig1}). In 2019, prior to the COVID-19 pandemic, work patterns are more office oriented, with 42.0\% of working days spent at the office and 35.3\% spent at home. In 2022, reflecting the widespread adoption of remote work practices, the proportion of office-based workdays more than halved to 20.7\%, while the share of home-based workdays rose to 55.3\%. While there was clear evidence of a return-to-office movement by 2023, the recovery is far from reverting to the previous normal. Office attendance increased to just 29.1\%, still well below the pre-pandemic benchmark, while home-based workdays remained elevated at 47.9\%, more than double the pre-pandemic level. Taken together, these patterns indicate that working from home has become a structural shift in daily work arrangements, one that the end of the pandemic has not fully undone despite some rebound in office presence during 2023.

\begin{figure}[H]
\centering
\includegraphics[width=0.9\textwidth]{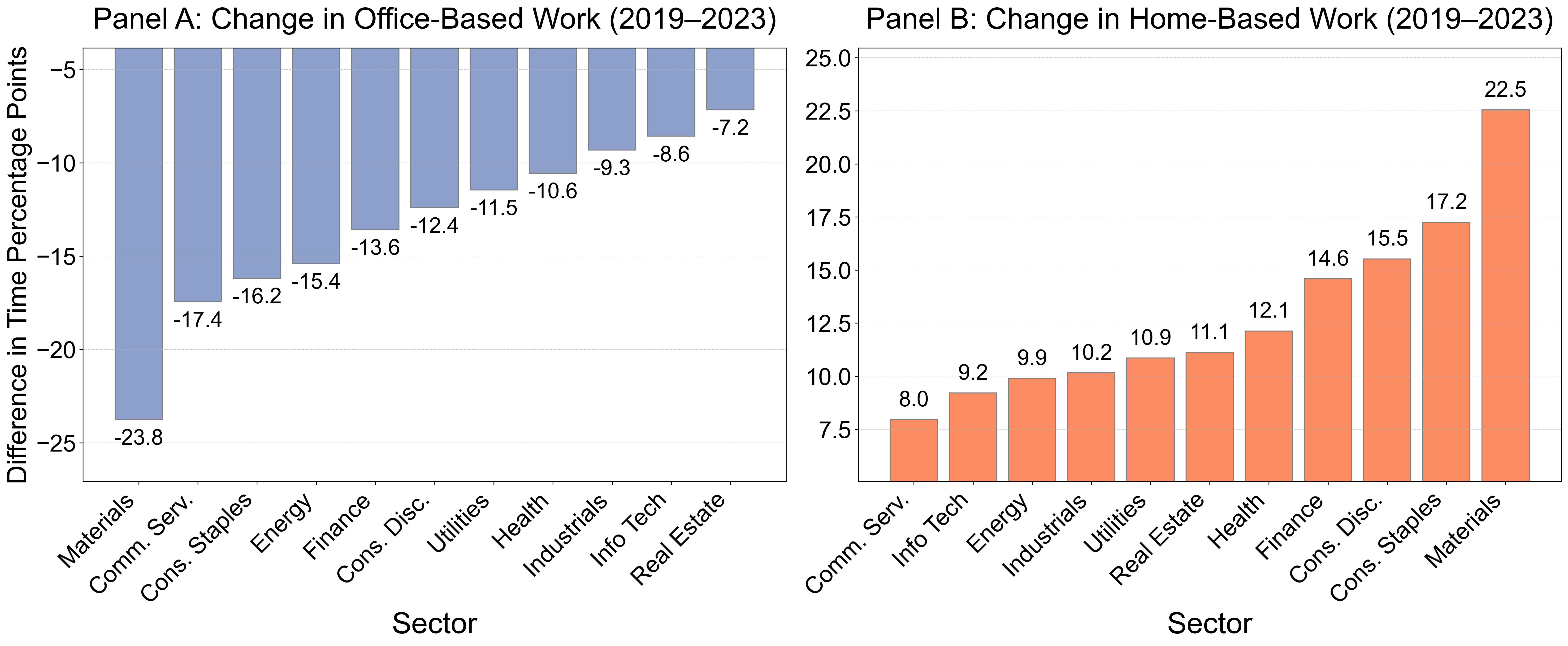}
\caption{\textbf{Sectoral Shifts in Workplace Attendance:} Office vs. Home‑Based Work Patterns Before and After the Pandemic}\label{fig2}
\end{figure}

Figure \ref{fig2} examines sector-level changes in workplace attendance between 2019 and 2023. Across all sectors, office-based workdays declined meaningfully over this period, though the magnitude of the decline varies considerably across industries. The largest reduction occurred in Materials, which saw a decline of 23.8 percentage points. Substantial drops are also observed in Communication Services (17.4 percentage points), Consumer Staples (16.2 percentage points), and Energy (15.4 percentage points).Sectors in the middle of the distribution, including Finance, Consumer Discretionary, and Utilities, clustered together with moderate declines ranging from 11.5 to 13.6 percentage points. Even sectors with comparatively smaller shifts, such as Information Technology (9.3 percentage points) and Real Estate (7.2 percentage points) experienced notable reductions in office attendance. Overall, the widespread negative shifts underscore that the move away from office-based work was not isolated to a few industries, but rather a broad-based adjustment affecting nearly every major sector.

Parallel to the decline in office presence, home-based workdays increased across all sectors between 2019 and 2023. The Materials sector exhibited the largest rise, with a 22.5 percentage point increase in time spent working from home. Consumer Staples (17.2 percentage points), Consumer Discretionary (15.5 percentage points), and Finance (14.6 percentage points) also experienced substantial increases. For industries like Materials and Consumer Staples, the magnitude of the increase in home-based work nearly mirrors the decline in office attendance, suggesting a relatively direct substitution of work location. By comparison, sectors such as Communication Services (8.0 percentage points) and Information Technology (9.2 percentage points) show more modest changes. While the sectors with the steepest declines in office-based work generally saw the largest gains in home-based work, the relationship is not perfectly symmetric, highlighting the presence of sector-specific operational constraints, task structures, and remote work feasibility differences.

In Appendix \ref{secA1} (Figures \ref{figA6} to \ref{figA10}), we analyze temporal shifts in work location behavior across five major metropolitan areas Chicago, Los Angeles, Miami, New York–New Jersey, and San Francisco, by examining the proportion of workdays spent in the office and at home in 2019, 2022, and 2023. While all regions exhibit a decline in office attendance and a corresponding rise in home-based work following the COVID-19 pandemic, the magnitude and persistence of these shifts vary significantly by metropolitan area. The New York area exhibits the most pronounced and persistent reallocation away from the office, while the Miami MSA shows the smallest changes, reflecting both regional policy responses and differences in sectoral composition, commuting costs, and work culture.

\begin{table}[H]
\centering
\captionsetup{justification=justified, width=\textwidth}
\begin{tabular}{lccc}
\hline
  & \textbf{2019} & \textbf{2022} & \textbf{2023} \\
\hline
Office & 442.1 & 427.3 & 406.5 \\

Home & 25.0 & 47.9 & 43.9 \\
\hline
\end{tabular}
\caption{Average number of minutes spent by people when working between 9AM to 5PM at home and office.}
\label{tab:working_minutes}
\end{table}

To understand how work location preferences and behaviors have evolved over time, we analyze the average time individuals spend at the office and home during standard working hours (9am to 5pm) across three years: 2019 (pre-pandemic), 2022 (pandemic adjustment phase), and 2023 (post-pandemic transition). Table \ref{tab:working_minutes} shows the results.

In 2019, prior to the COVID-19 pandemic, the average individual spent approximately 442.1 minutes or over 92\% of the 480-minute workday at the office, with only 25.0 minutes spent at home. This reflects a predominantly office-centered work culture, consistent with traditional expectations around full-time in-person work. 

By 2022, a clear shift emerged. Average time spent in the office fell modestly to 427.3 minutes—about 15 minutes below the 2019 level—while home-based presence during work hours nearly doubled to 47.9 minutes. In 2023, the pattern continued, though more gradually. Office time declined further to 406.5 minutes, and home time remained elevated at 43.9 minutes, well above the 2019 baseline. This persistence is notable given that 2023 is widely regarded as the year when employers pushed most assertively for a return to the office. Taken together, these dynamics indicate that a pattern initially viewed as a pandemic-era anomaly has instead evolved into a structural shift in work practices.

Taken together, Table 1 and Figure 1 reveal a subtle but important shift: office attendance has declined not only in frequency but also in intensity, as the time spent in the office on a given day continues to shrink. These patterns support the view that the COVID-19 pandemic triggered a lasting transformation in how work is distributed across space and time, rather than a temporary disruption.

\subsection{A new ``midweek mountain''}
\begin{figure}[H]
\centering
\includegraphics[width=0.9\textwidth]{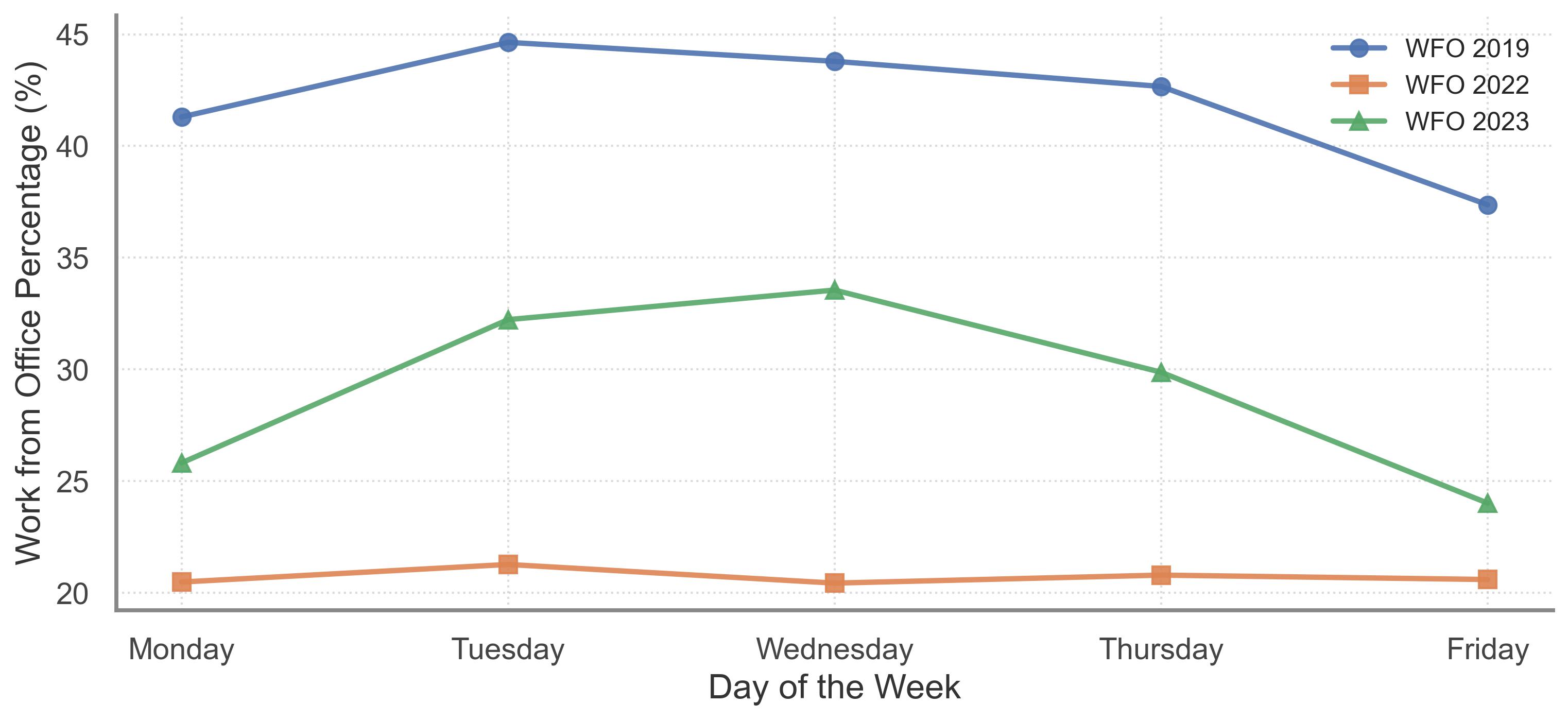}
\caption{\textbf{Weekday Work‑from‑Office Proportions in 2019, 2022, and 2023} 
}\label{fig3}
\end{figure}

In Figure \ref{fig3}, we examine the variation in work-from-office (WFO) percentages by day of the week across the years 2019, 2022, and 2023. In 2019, work-from-office patterns exhibited an end-of-week decline. Office attendance was similar from Monday to Thursday (41\% to 43\%) and clearly declined on Friday (37\%). Year 2022 sees a substantial and persistent reduction in office attendance across all days of the week. The WFO percentage remains nearly flat from Monday to Friday, fluctuating narrowly around 20\%. The uniformly low levels of office attendance suggest that remote work practices remained largely in place a year out of the pandemic.

By 2023, a recovery in office attendance is evident, and significantly more profound in the middle of the week. WFO shares rise from comparatively low levels on Monday (26\%) to a clear peak on Wednesday (nearly 34\%), before declining again on Thursday and falling sharply on Friday (24\%). Relative to pre-pandemic, this “midweek mountain” is a new phenomenon: Tuesdays and Wednesdays stand out as the dominant in-office days, whereas Mondays and Fridays are disproportionately remote.

These evolving day-of-week patterns suggest that firms and workers are converging toward hybrid arrangements that compress in-person work into the middle of the week. The steep midweek mountain in 2023 may well sustain as a longer run trend toward more flexible workweeks, in which employees coordinate face-to-face activities on a few core days and reserve the beginning and end of the week for remote work. Beyond spatial reorganization, continuing advances in digital tools and AI-enhanced productivity also raise the prospect that similar levels of output can be achieved with fewer labor hours, making this paradigm shift potentially conducive to a shorter workweek more generally.


\subsection{Where do people spent their time during workday?}

\begin{figure}[H]
\centering
\includegraphics[width=1\textwidth]{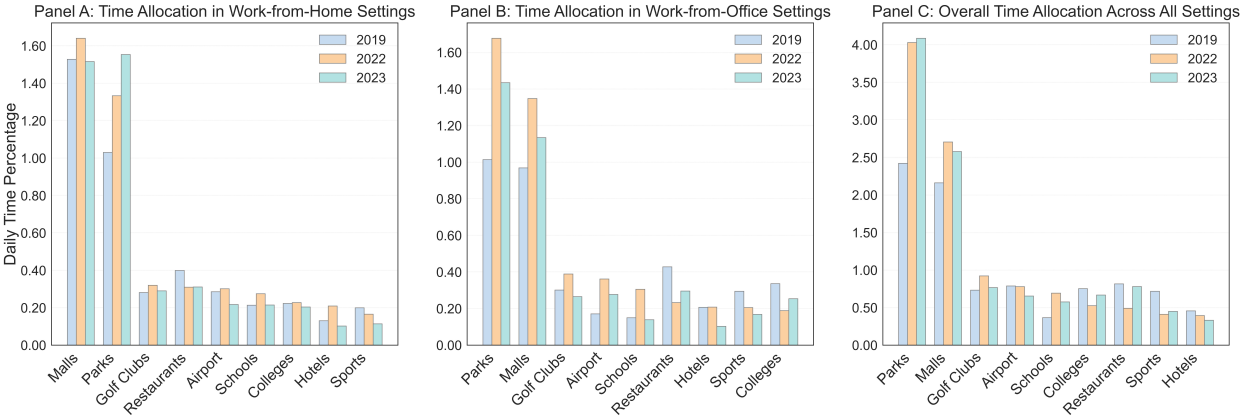}
\caption{\textbf{Daily Time Allocation at Key Points of Interest On Work‑from‑Home vs. Work‑from‑Office Days vs. All Working Days, 2019–2023} 
}\label{fig4}
\end{figure}

We next examine how people spend their time at different types of locations during work days in Figure \ref{fig4}.\footnote{We exclude the category of non-home residential properties from the analysis in this Figure as it is challenging to determine whether these locations are mainly used for work or leisure activities.} This analysis is only possible with detailed geolocation data as in our study, as survey responses typically do not contain such granularity. When working from home, individuals in 2022 spent more time across nearly all categories of public or semi-public places, e.g., malls, parks, and golf clubs, compared to 2019 with the exceptions of restaurants, colleges and sports where the difference is marginal. This pattern suggests that during the height of remote work adoption in 2022, individuals exhibited greater mobility throughout the workday. Rather than strictly adhering to indoor work routines, many used the autonomy brought by Covid-era remote work for more short breaks or errands, reflecting a more relaxed or hybrid interpretation of home-based work.

What is more surprising is that mobility also increases even more during post-pandemic period when people work from office. Office days become less “office bound” than before the pandemic. Employees increasingly combine time at the workplace with trips to nearby amenities or other locations, suggesting that the return to the office has been accompanied by more flexible and mobile daily routines rather than a simple reversion to pre-pandemic behavior.\footnote{The results are qualitatively similar if we adopt alternative definitions of work-from-office days (e.g., Figure \ref{figrobust} in the Appendix).} Overall, on the extensive margin, the total time spent at many public and semi-public points of interest during work hours increases nearly 14\%  after the pandemic and remains elevated in 2023 (Panel C of Figure \ref{fig4}).


Appendix \ref{secA1} (Figures \ref{figA1} to \ref{figA5}) further analyzes time spent at various locations in the different MSA areas. Time allocation across different types of locations exhibits meaningful variation across metropolitan areas. For example, malls and parks become much more popular locations to visit after the pandemic for San Francisco, New York, and Miami. Coastal cities such as San Francisco and Miami also show a marked increase in time spent at port-related and hydro-infrastructure locations. The geography of daily work/life continues to be determined as much by local culture and infrastructure as by economic forces.



\begin{figure}[H]
\centering
\includegraphics[width=1\textwidth]{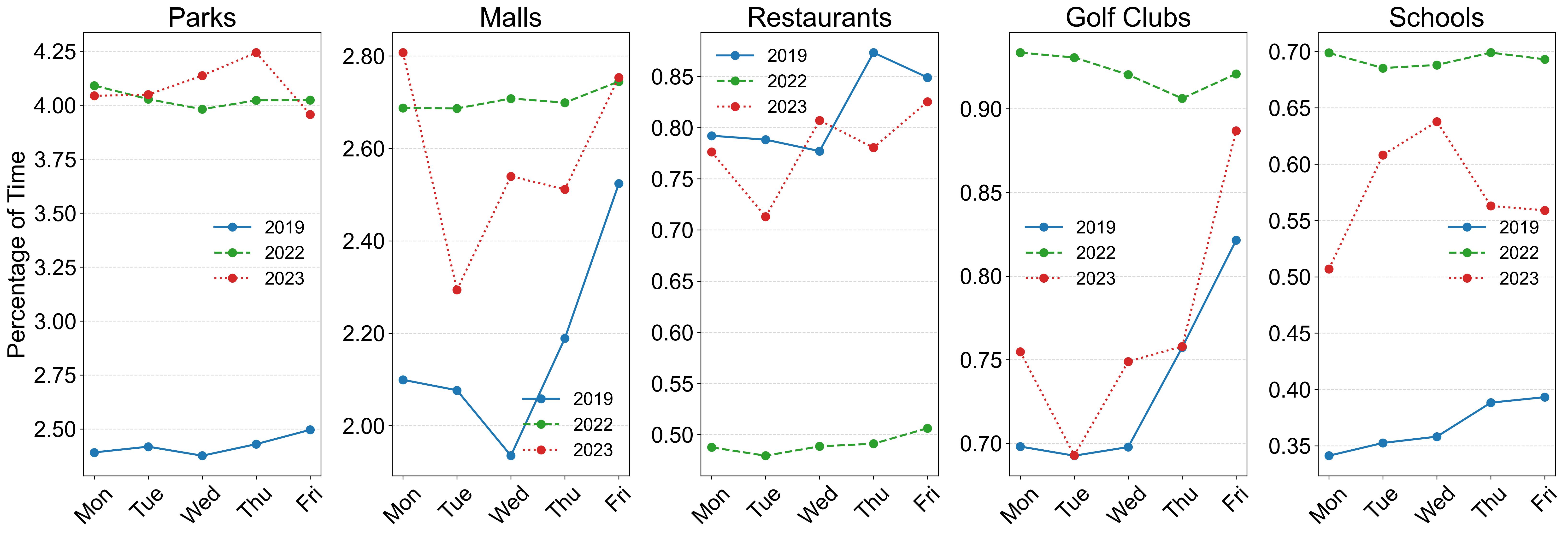}
\caption{\textbf{Weekday Dynamics of Time Allocation across Location Categories from 2019 to 2023}
}\label{fig5}
\end{figure}

In Figure \ref{fig5}, we examine the distributions of time spent at different location categories, including parks, malls, restaurants, golf clubs, and schools, across the days of the week, from 2019 to 2023. Some key patterns emerge. First, there is an elevated tendency to visit all of these types of locations in 2022, with a relatively flat weekday visit pattern. Second, while visits to some categories return nearly to their 2019 patterns, some others do not. For example, visits to parks and malls stay as high as their 2022 levels throughout the work week in 2023. Third, the midweek pattern becomes more dramatic in 2023 as workers tend to visit non-work locations much more on Mondays and Fridays. Overall, the patterns again suggest a permanent shift in people's time allocation in a weekly cycle.


\section{Conclusion}\label{sec13}

Using billions of geolocation pings, we show that the pandemic-induced shift in work patterns represents a durable structural change rather than a temporary disruption. Even after a partial recovery in 2023, office attendance remains well below its pre-pandemic level. The emergence of a coordinated, midweek-focused schedule indicates that firms and employees are converging on stable hybrid routines. Together with digital tools and AI-enhanced productivity, this reorganization is conducive to a shortened workweek, as similar output can increasingly be achieved with fewer on-site days and fewer total work hours. We also document a decline in both the frequency and the intensity of office work.

The data further reveal a blurring of boundaries between work and non-work activities. Increased time spent at parks and malls during remote workdays indicates that flexibility is reshaping daily routines rather than merely shifting work from office to home. For firms, these patterns call for policies that accommodate stable hybrid scheduling and more concentrated in-person work. For urban planners and policymakers, they signal lasting changes in mobility patterns and the economic role of central business districts. More broadly, the future of work now turns not only on where work takes place, but on how work time itself is organized under sustained technological change.





\newpage

\appendix

\begin{center}
\Large {Supplementary Appendix to ``Mapping the Midweek Mountain:  The New Geography of Hybrid Work''}
\end{center}

\section{Time Allocation Patterns Across Different MSAs}\label{secA1}

\begin{figure}[H]
\centering
\includegraphics[width=1\textwidth]{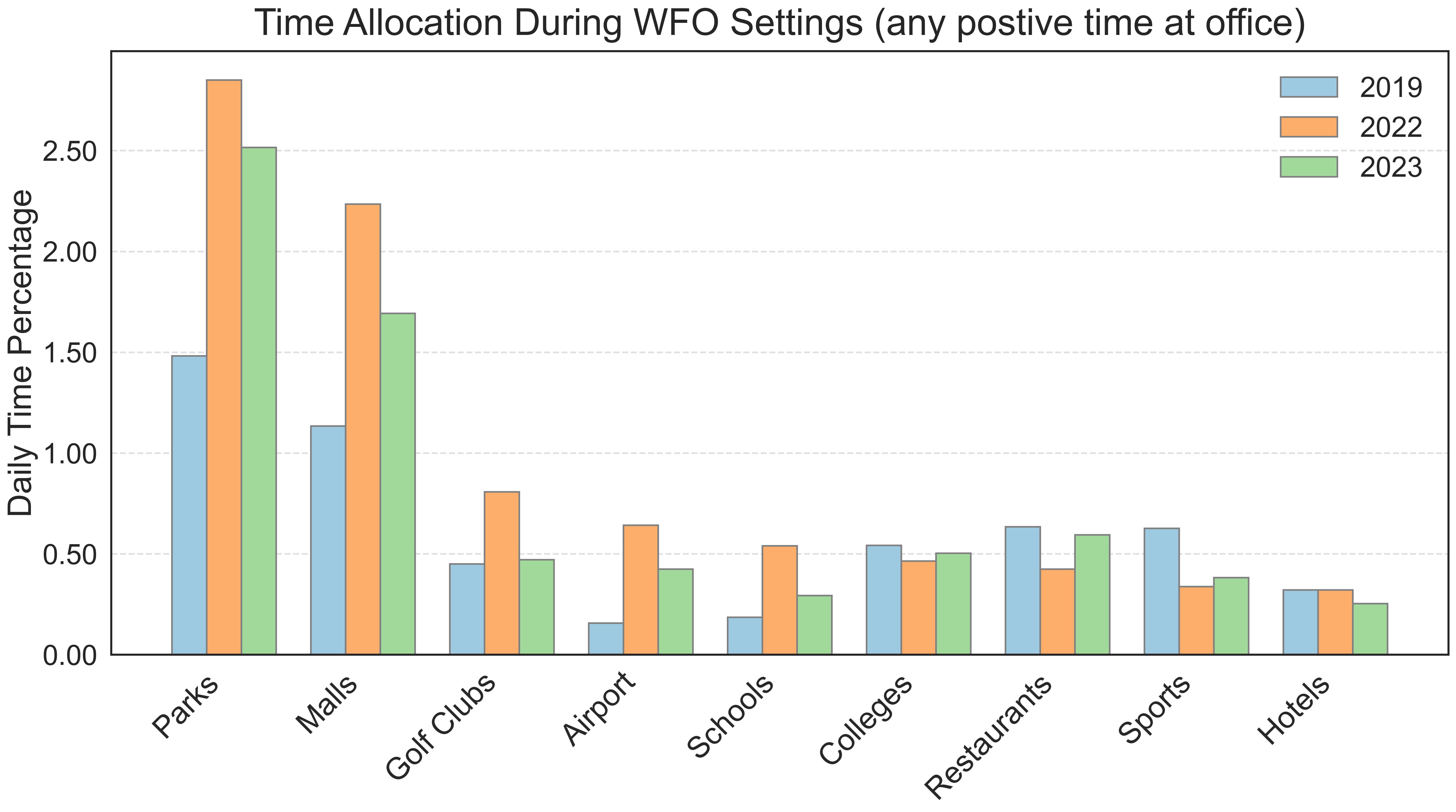}
\caption{\textbf{Daily Time Allocation at Key Points of Interest During Work-from-Office Days.}
A person-day is classified as WFO if the individual spends any positive amount of time at the office. This definition ensures that the results are not sensitive to alternative threshold choices for identifying WFO days.}\label{figrobust}
\end{figure}

\begin{figure}[H]
\centering
\includegraphics[width=0.9\textwidth]{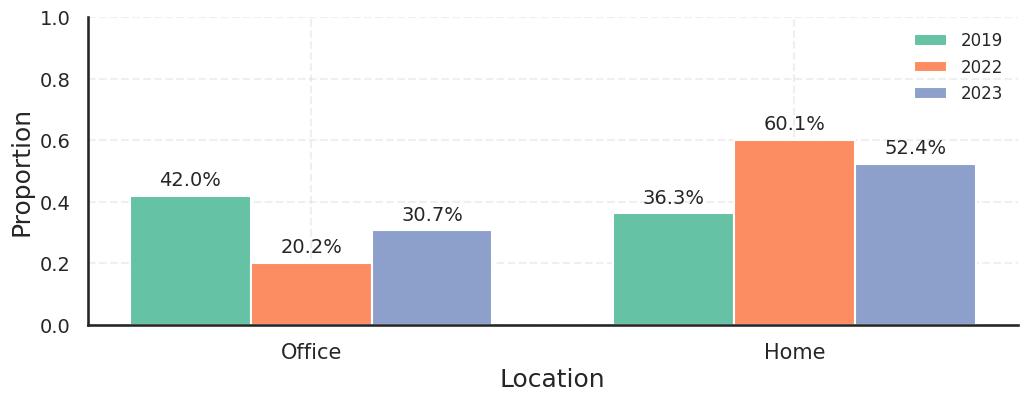}
\caption{\textbf{Shifts in Work Location Preferences:} Proportion of Workdays Spent in the Office vs. at Home in 2019, 2022, and 2023 in the Chicago MSA}\label{figA6}
\end{figure}

\begin{figure}[H]
\centering
\includegraphics[width=0.9\textwidth]{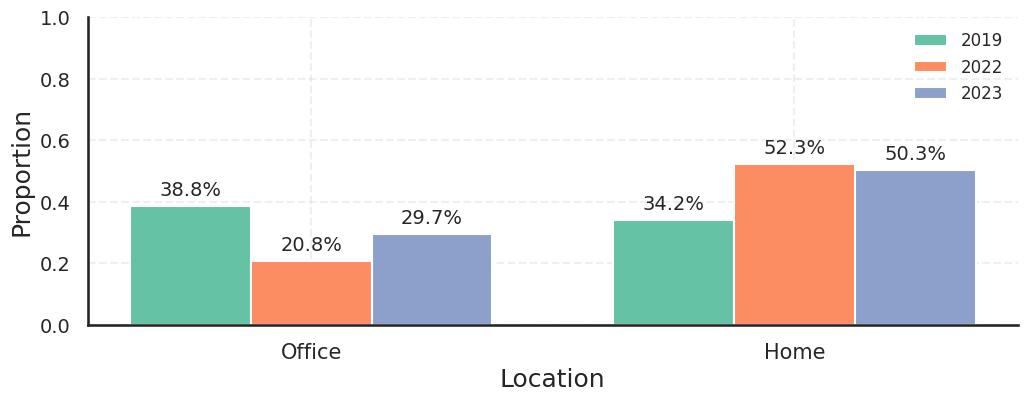}
\caption{\textbf{Shifts in Work Location Preferences:} Proportion of Workdays Spent in the Office vs. at Home in 2019, 2022, and 2023 in the Los Angeles MSA}\label{figA7}
\end{figure}

\begin{figure}[H]
\centering
\includegraphics[width=0.9\textwidth]{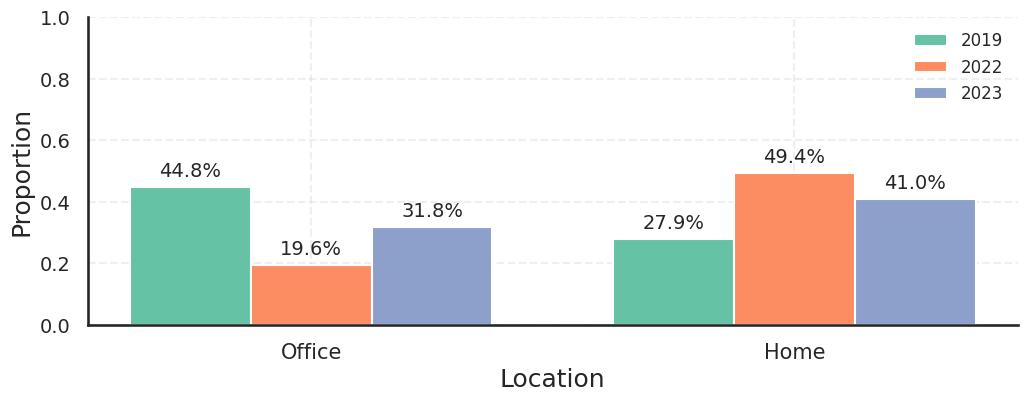}
\caption{\textbf{Shifts in Work Location Preferences:} Proportion of Workdays Spent in the Office vs. at Home in 2019, 2022, and 2023 in the Miami MSA}\label{figA8}
\end{figure}

\begin{figure}[H]
\centering
\includegraphics[width=0.9\textwidth]{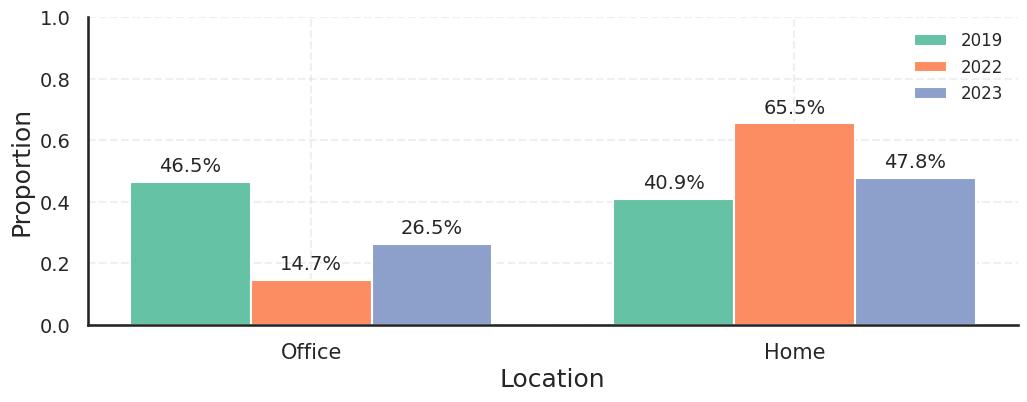}
\caption{\textbf{Shifts in Work Location Preferences:} Proportion of Workdays Spent in the Office vs. at Home in 2019, 2022, and 2023 in the New York MSA}\label{figA9}
\end{figure}

\begin{figure}[H]
\centering
\includegraphics[width=0.9\textwidth]{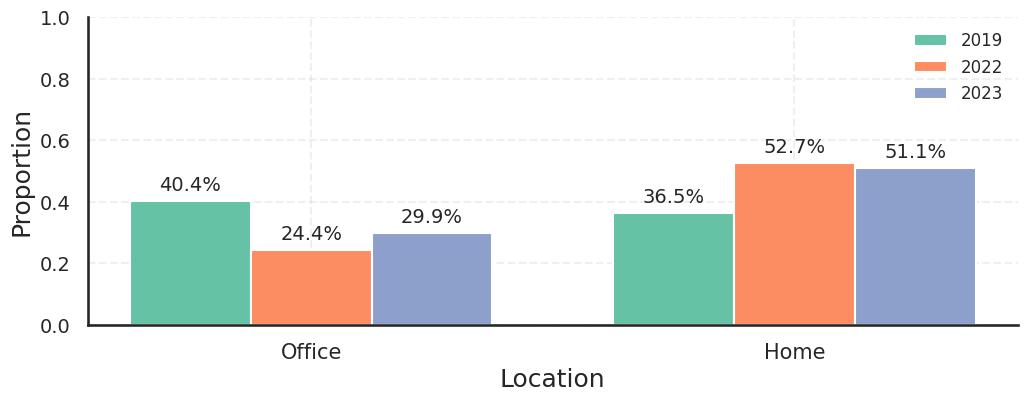}
\caption{\textbf{Shifts in Work Location Preferences:} Proportion of Workdays Spent in the Office vs. at Home in 2019, 2022, and 2023 in the San Francisco MSA}\label{figA10}
\end{figure}

\begin{figure}[H]
\centering
\includegraphics[width=1\textwidth]{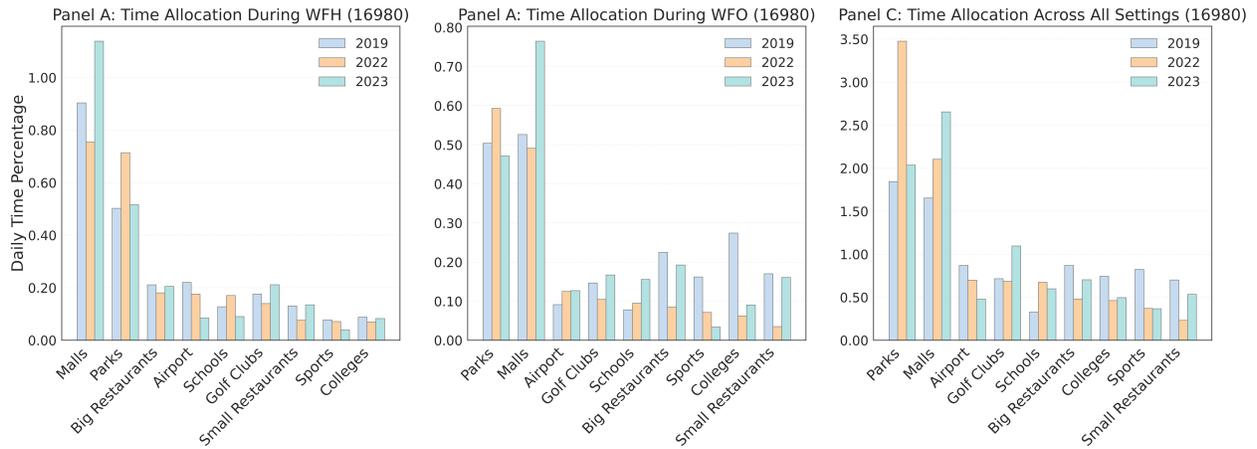}
\caption{\textbf{Daily Time Allocation at Key Points of Interest During Work‑from‑Home vs. Work‑from‑Office, 2019–2023 at the Chicago MSA}, Highlights how average minutes per day spent at apartments, malls, parks, etc., shifted across three years under home and office work settings}\label{figA1}
\end{figure}

\begin{figure}[H]
\centering
\includegraphics[width=1\textwidth]{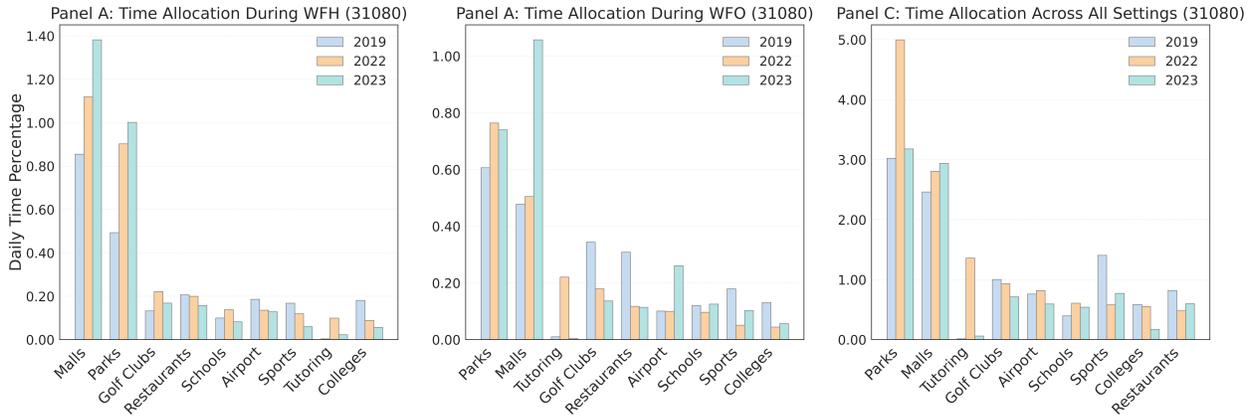}
\caption{\textbf{Daily Time Allocation at Key Points of Interest During Work‑from‑Home vs. Work‑from‑Office, 2019–2023 at the Los Angeles MSA}, Highlights how average minutes per day spent at apartments, malls, parks, etc., shifted across three years under home and office work settings}\label{figA2}
\end{figure}

\begin{figure}[H]
\centering
\includegraphics[width=1\textwidth]{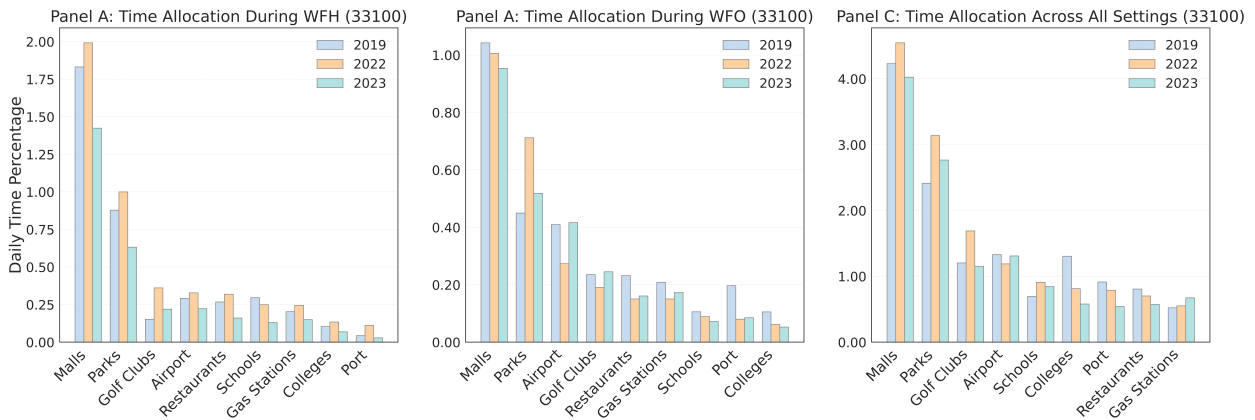}
\caption{\textbf{Daily Time Allocation at Key Points of Interest During Work‑from‑Home vs. Work‑from‑Office, 2019–2023 at the Miami MSA}, Highlights how average minutes per day spent at apartments, malls, parks, etc., shifted across three years under home and office work settings}\label{figA3}
\end{figure}

\begin{figure}[H]
\centering
\includegraphics[width=1\textwidth]{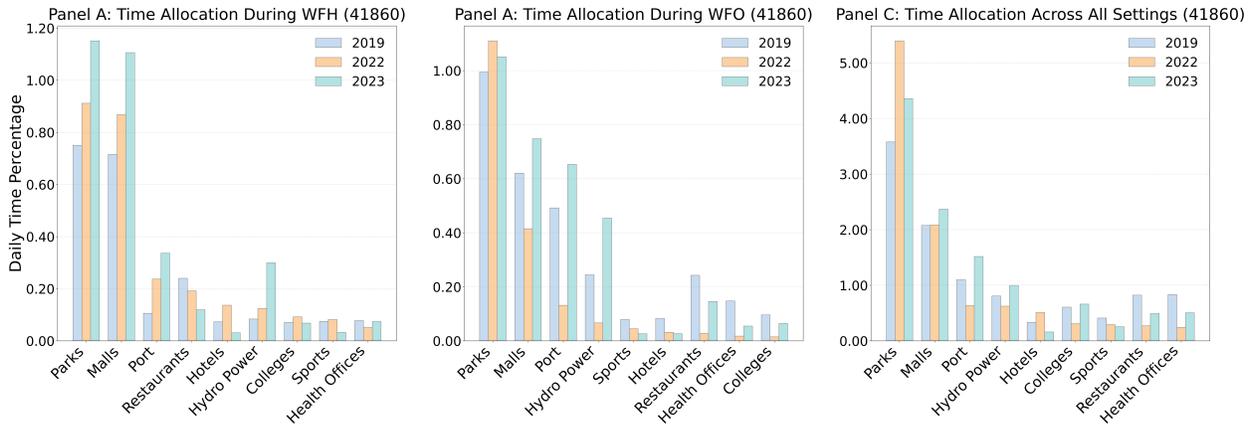}
\caption{\textbf{Daily Time Allocation at Key Points of Interest During Work‑from‑Home vs. Work‑from‑Office, 2019–2023 at the San Francisco MSA}, Highlights how average minutes per day spent at apartments, malls, parks, etc., shifted across three years under home and office work settings}\label{figA4}
\end{figure}

\begin{figure}[H]
\centering
\includegraphics[width=1\textwidth]{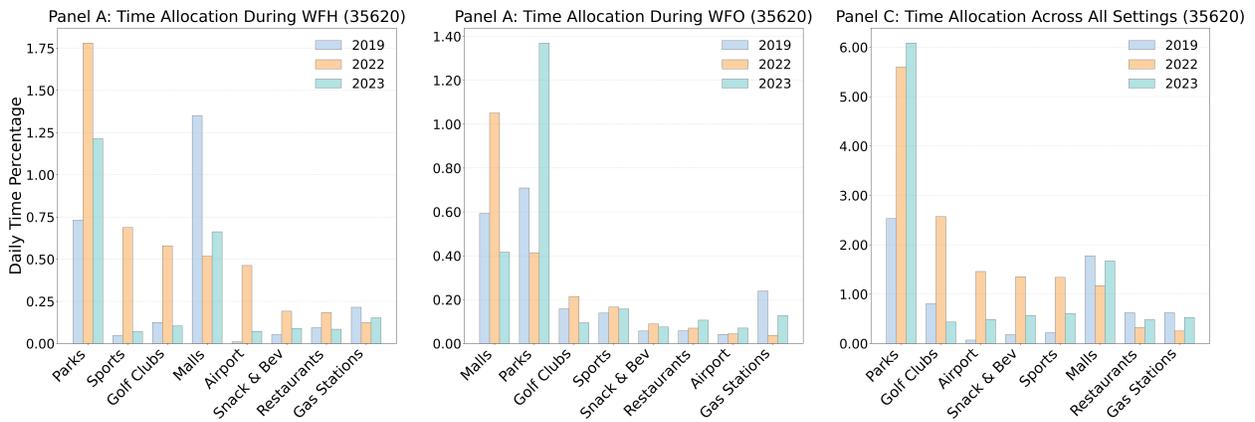}
\caption{\textbf{Daily Time Allocation at Key Points of Interest During Work‑from‑Home vs. Work‑from‑Office, 2019–2023 at the New York MSA}, Highlights how average minutes per day spent at apartments, malls, parks, etc., shifted across three years under home and office work settings}\label{figA5}
\end{figure}


\begin{thebibliography}{99}

\bibitem[Agrawal, Gans \& Goldfarb(2019)]{agrawal2019}
Agrawal, A., Gans, J. S. \& Goldfarb, A. Artificial intelligence: The ambiguous labor market impact of automating prediction. \textit{J. Econ. Perspect.} \textbf{33}, 31–50 (2019).

\bibitem[Aksoy, Barrero, Bloom, Davis, Dolls \& Zarate(2022)]{aksoy2022}
Aksoy, C. G., Barrero, J. M., Bloom, N., Davis, S. J., Dolls, M. \& Zarate, P. Working from home around the world. \textit{Brookings Papers on Economic Activity} \textbf{2022}, 259–331 (2022).

\bibitem[Allen, Golden \& Shockley(2015)]{allen2015}
Allen, T. D., Golden, T. D. \& Shockley, K. M. How effective is telecommuting? Assessing the status of our scientific findings. \textit{Psychol. Sci. Public Interest} \textbf{16}, 40–68 (2015).

\bibitem[Autor, Dorn \& Hanson(2022)]{autor2022}
Autor, D., Dorn, D. \& Hanson, G. H. When work disappears: Manufacturing decline and the falling marriage market value of young men. \textit{American Economic Review: Insights} \textbf{4}, 161–178 (2022).

\bibitem[Bailey \& Kurland(2002)]{bailey2002}
Bailey, D. E. \& Kurland, N. B. A review of telework research: Findings, new directions, and lessons for the study of modern work. \textit{J. Organ. Behav.} \textbf{23}, 383–400 (2002).

\bibitem[Barrero, Bloom \& Davis(2021)]{barrero2021}
Barrero, J. M., Bloom, N. \& Davis, S. J. Why working from home will stick. \textit{Am. Econ. Rev.} \textbf{111}, 1756–1790 (2021).

\bibitem[Barrero, Bloom \& Davis(2023)]{barrero2023}
Barrero, J. M., Bloom, N. \& Davis, S. J. The evolution of work from home. \textit{J. Econ. Perspect.} \textbf{37}, 23–50 (2023).

\bibitem[Bloom, Han \& Liang(2024)]{bloom2024}
Bloom, N., Han, R. \& Liang, J. Hybrid working from home improves retention without damaging performance. \textit{Nature} \textbf{630}, 920–925 (2024).

\bibitem[Brynjolfsson(2014)]{brynjolfsson2014}
Brynjolfsson, E. \textit{The Second Machine Age: Work, Progress, and Prosperity in a Time of Brilliant Technologies}. WW Norton Company (2014).


\bibitem[Brynjolfsson, Horton, Ozimek, Rock, Sharma \& TuYe(2020)]{brynjolfsson2020}
Brynjolfsson, E., Horton, J. J., Ozimek, A., Rock, D., Sharma, G. \& TuYe, H. Y. COVID-19 and remote work: An early look at US data. \textit{NBER Working Paper} No. 27344 (2020).

\bibitem[Cooper \& Kurland(2002)]{cooper2002}
Cooper, C. D. \& Kurland, N. B. Telecommuting, professional isolation, and employee development in public and private organizations. \textit{J. Organ. Behav.} \textbf{23}, 511–532 (2002).

\bibitem[Choudhury, Khanna, Makridis \& Schirmann(2024)]{choudhury2024}
Choudhury, P., Khanna, T., Makridis, C. A. \& Schirmann, K. Is hybrid work the best of both worlds? Evidence from a field experiment. \textit{Rev. Econ. Stat.}, 1–24 (2024).

\bibitem[Dingel \& Neiman(2020)]{dingel2020}
Dingel, J. I. \& Neiman, B. How many jobs can be done at home? \textit{J. Public Econ.} \textbf{189}, 104235 (2020).

\bibitem[Gajendran \& Harrison(2007)]{gajendran2007}
Gajendran, R. S. \& Harrison, D. A. The good, the bad, and the unknown about telecommuting: Meta-analysis of psychological mediators and individual consequences. \textit{J. Appl. Psychol.} \textbf{92}, 1524–1541 (2007).

\bibitem[Gallup(2023)]{gallup2023}
Gallup. The future of the office has arrived: It's hybrid. (2023). Available at: \url{https://www.gallup.com/workplace/511994/future-office-arrived-hybrid.aspx}.

\bibitem[Gibbs, Mengel \& Siemroth(2024)]{gibbs2024}
Gibbs, M., Mengel, F. \& Siemroth, C. Employee innovation during office work, work from home and hybrid work. \textit{Sci. Rep.} \textbf{14}, 17117 (2024).

\bibitem[Golden, Veiga \& Dino(2008)]{golden2008}
Golden, T. D., Veiga, J. F. \& Dino, R. N. The impact of professional isolation on teleworker job performance and turnover intentions: Does time spent teleworking, interacting face-to-face, or having access to communication-enhancing technology matter? \textit{J. Appl. Psychol.} \textbf{93}, 1412–1421 (2008).

\bibitem[Howard, Liebersohn \& Ozimek(2023)]{howard2023}
Howard, G., Liebersohn, J. \& Ozimek, A. The short- and long-run effects of remote work on U.S. housing markets. \textit{J. Financ. Econ.} \textbf{150}, 166–184 (2023).

\bibitem[Hsu, Liu, Nguyen, Chien \& Mostafavi(2024)]{hsu2024}
Hsu, C.-W., Liu, C., Nguyen, K. M., Chien, Y.-H. \& Mostafavi, A.
Do human mobility network analyses produced from different location-based data sources yield similar results across scales.
\textit{Computers, Environment and Urban Systems} \textbf{107}, 102052 (2024).

\bibitem[Kastle(2023)]{kastle2023}
Kastle Systems. Back to work barometer. (2023). Available at: \url{https://www.kastle.com/safety-wellness/getting-america-back-to-work/}.

\bibitem[Kniffin, Narayanan, Anseel, Antonakis, Ashford, Bakker, Bamberger, Bapuji, Bhave, Bisqueret, Bitektine \& Bono(2021)]{kniffin2021}
Kniffin, K. M. \textit{et al.} COVID-19 and the workplace: Implications, issues, and insights for future research and action. \textit{Am. Psychol.} \textbf{76}, 63–77 (2021).

\bibitem[Mondragon \& Wieland(2022)]{mondragon2022}
Mondragon, J. A. \& Wieland, J. Housing demand and remote work. \textit{Am. Econ. Rev.} \textbf{112}, 3171–3204 (2022).

\bibitem[Miyauchi, Nakajima \& Redding(2021)]{miyauchi2021}
Miyauchi, Y., Nakajima, K. \& Redding, S. J. The economics of spatial mobility: Theory and evidence using smartphone data. \textit{NBER Working Paper} No. 28497 (2021).

\bibitem[Waber, Magnolfi \& Lindsay(2014)]{waber2014}
Waber, B., Magnolfi, J. \& Lindsay, G. Workspaces that move people. \textit{Harvard Business Review} \textbf{92}, 68–77 (2014).

\bibitem[Wirz(2025)]{wirz2025}
Wirz, M. The 'zombie buildings' at the heart of the office meltdown. \textit{Wall Street Journal} (27 Apr. 2025). Available at: \url{https://www.wsj.com/real-estate/commercial/chicago-office-buildings-real-estate-market-1913fe3a} [Accessed 5 May 2025].

\bibitem[Yang, Holtz, Jaffe, Suri, Sinha, Weston, Joyce, Shah, Sherman, Hecht \& Teevan(2022)]{yang2022}
Yang, L., Holtz, D., Jaffe, S., Suri, S., Sinha, S., Weston, J., Joyce, C., Shah, N., Sherman, K., Hecht, B. \& Teevan, J. The effects of remote work on collaboration among information workers. \textit{Nature Human Behaviour} \textbf{6}, 43–54 (2022).

\end{thebibliography}
\end{document}